\documentstyle[12pt,epsf]{article}
\begin{document}
\begin{center}
{\large\sf
RELATIVISTIC EFFECTS IN TWO-BODY %\\[5mm]
SYSTEMS: \\[3mm]
$\pi$- AND $K$- MESONS AND DEUTERON%
\footnote{Poster presented by V.E.T.\ at the
15th International
Conference on Few-Body Problems in Physics, Groningen,
July 22-26, 1997.} }\\[8mm]
{\sf A.F.~KRUTOV}\\
{\sf Samara State University, 443011, Samara,
Russia}\\[5mm]
\begin{tabular}{c}
{\sf \underline{V.E.~TROITSKY}}\\
\end{tabular}\\
{\sf
Nuclear Physics Institute MSU, 119899, Moscow, Russia}\\[10mm]
\end{center}
\begin{abstract}
The electromagnetic form factors of $\pi$ and $K$ mesons and
deuteron are calculated in modified impulse approximation using
instant form of relativistic Hamiltonian dynamics.
The different model wave functions are used. Meson wave function
parameters are fixed by fitting the mean square radius of meson.
The internal quark structure is taken into account through
electromagnetic quark form factor and quark anomalous magnetic
moments. Results of our calculations of electroweak
structure of pion and kaon and electromagnetic deuteron
properties agree well with the available experimental data.
The meson form factors asymptotics at large momentum transfer
is the same as in perturbative QCD. Some predictions about
CEBAF experiments are given.

\end{abstract}

%
%{\LARGE\sf
In recent years the interest
has been renewed  to the study of
electromagnetic form factors of composite systems:
of pseudoscalar mesons and,
particularly, $\pi$- and $K$- mesons, and of the deuteron.
This fact is due, first of all, to the
experiments at CEBAF concerning the measurement
of pion, kaon  and deuteron form factors.
It is possible that these  experiments
will enable to choose between different theoretical
models whose predictions differ rather strongly.
\smallskip

Such a difference of theoretical results seems to be
quite natural, because one encounters a lot of
difficulties while calculating the structure of composite
particles.
The main difficulties are the following two.
First, the importance of relativistic effects
in the whole range of momentum transfer.
Second, the problem of calculation of the "soft"
structure which cannot be obtained
from perturbative QCD.
The soft part, which describes the structure at
long and intermediate distances, needs nonperturbative
approaches. (A controversy still exists concerning a scale of
momentum transfers characteristic for the boundary between the
nonperturbative to perturbative regime.)

The relativistic Hamiltonian dynamics (RHD) (for
a review see \cite{KeP91})
is one of such nonperturbative ways.
RHD unifies
potential approach to composite systems and the condition of
Poincar\'e-invariance.  This  method is based on the direct
realization of the Poincar\'e group algebra or, in other words,
of the relativistic invariance condition in the few body Hilbert space.
RHD can be formulated in different ways ( different relativistic dynamics).
At present the light front dynamics is the most popular one.
\smallskip

Here we present a
relativistic treatment of the problem of soft electromagnetic
structure in the framework of alternative form of relativistic
dynamics, the so called instant form (IF) of RHD.  Our approach,
which describes well the available data for the
elastic form factors for the charged $\pi$ (see figs.1 -- 3) and
$K$ (see fig.4) mesons and deu\-te\-ron (see figs.5 -- 7), will
be briefly outlined here.  The details are partially given
in \cite{KrT96}.  The IF form of relativistic dynamics,
although not widely used, has some advantages. The calculations
can be performed in a natural straightforward way without
special coordinates.  IF is particularly convenient to discuss
the nonrelativistic limit of relativistic results. The space
reflection and time reversal operators do not depend on the
interaction (contrary to the case of light front dynamics)
\cite{Lev95}.  This approach is obviously rotationally invariant
so that IF is the most suitable for spin and polarization
problems.  Our method gives the same form factors asymptotics as
in QCD. It is worth to notice that the relativization is
performed in the same way for quark--antiquark systems as for
nucleon--nucleon composite systems. At last, for the systems
with spin 1 (e.g., deuteron and $\rho$  -- meson) the solution
for the formfactors is unique (contrary to the case of light
front dynamics).
\smallskip

In our
approach it is possible  to construct the current operator of
composite system which satisfies the conservation law and the
relativistic covariance condition.  The calculation is based on
the representation of two-body current matrix element in terms
of the free two-particle form factors $g(s,Q^2,s')$ where $s,s'$
are invariant masses of free two-particle system and $Q^2$ is
transfer momentum square.  $g(s,Q^2,s')$ is a matrix in orbital
moments $l,l'$ in the case of deuteron.  The final expression
for electromagnetic form factor has the following form:
\begin{equation}
F(Q^2) =
\int\,d\sqrt{s}\,\sqrt{s'}\,\varphi(s)\,g(s,Q^2,s')\,\varphi(s')
\label{result}
\end{equation}
Here $\varphi(s)$ is wave function which is calculated in RHD.
Results of our calculations of electroweak structure
of pion and kaon and electromagnetic deuteron properties
agree well with the available experimental data.
\smallskip

The following meson wave functions were utilized:

1. A gaussian or harmonic oscillator (HO)
wave function
\begin{equation}
u(k)= N_{HO}\,
\hbox{exp}\left(-{k^2}/{2b^2}\right).
\label{HO-wf}
\end{equation}
\noindent (curve 1 on figs.1,4;
for comparison the nonrelativistic limit of (\ref{result}) is
given by the curve 6 on figs.1,4).

2. A power-law (PL) wave function (see e.g. \cite{CaG95},
\cite{Sch94})
\begin{equation}
u(k) =N_{PL}\,{(k^2/b^2 +
1)^{-n}}\>,\quad n = 2\>,3\>.
\label{PL-wf}
\end{equation}
\noindent (curves 2,3 on figs.1,4).

3. The wave function with linear confinement from Ref.\cite{Tez91}:
\begin{equation}
u(r) = N_T \,\exp(-\alpha r^{3/2} - \beta r)\>,\quad \alpha =
\frac{2}{3}\sqrt{2\,M\,a}\>,\quad \beta = M\,b\>.
\label{Tez91-wf}
\end{equation}
$a\>,b$ -- parameters of linear and Coulomb parts of potential,
respectively.
\noindent (curve 4 on figs.1,4).

The values of the parameters are choosen so as to describe
experimental mean square radius and
lepton decay constant of mesons.

The internal quark structure can be described by the
introduction of electromagnetic quark form factor and anomalous
quark magnetic moment. Electromagnetic quark structure was
chosen in the form:
\begin{equation}
f_q(Q^2) = \frac{1}{1 +
\ln(1+ <r^2_q>Q^2/6)}\>.
\label{quark}
\end{equation}

$<r^2_q>$ -- is the mean square radius  of the quark.
The choice of form (\ref{quark}) does not violate the
asymptotics which takes place in our approach at
$Q^2\>\to\>\infty\>,\> M_q\>\to\>$0,
which in our approach is:  \begin{equation} F_c(Q^2)
\quad \sim\quad Q^{-2}\>.\label{Q-inf} \end{equation}

The model dependence of electromagnetic  pion form factor in
region of CEBAF experiments occurs to be comparatively weak.
On the other hand, pion
electromagnetic form factor depends strongly on the anomalous
quark magnetic moments.
Analogous results take place for kaon.
\smallskip

%\newpage

Our results lead to the following conclusions.
\smallskip

1. Most likely the CEBAF experiments on the measurements of
$\pi$-- and $K$-- mesons form factors at $Q^2\>\leq\>$3
(GeV/c)$^2$ will not allow for the possibility to extract the
best model of quark interaction in view of weak dependence on
the choice of wave function.

2. It seems reasonable to say that CEBAF
experiments will give the possibility to estimate the manifestation of
internal quark structure.

3. In the IF of RHD the meson form factors asymptotics at large
momentum transfer is the same as in perturbative QCD. The
asymptotics is now determined by relativistic kinematics only,
specifically by the relativistic effect of spin rotation, and
does not depend on the choice of the quark wave function, that
is of the quark interaction model.

4. IF of RHD describes well the experimental data on elastic
$ed$ -- scattering.

V.E.T. is indebted to the organizers of the 15th International
Conference on Few-Body Problems in Physics for the possibility
to attend the meeting and to present these results. This work is
supported in part by the Russian Foundation for Basic Research
(grant no.~96-02-17288).

%} %\newpage %\vfill %{\large\sf %\noindent

\begin{figure}[p]
\begin{picture}(18,11)(-1,0)
\put(4,1){\makebox[8cm]{\epsfbox{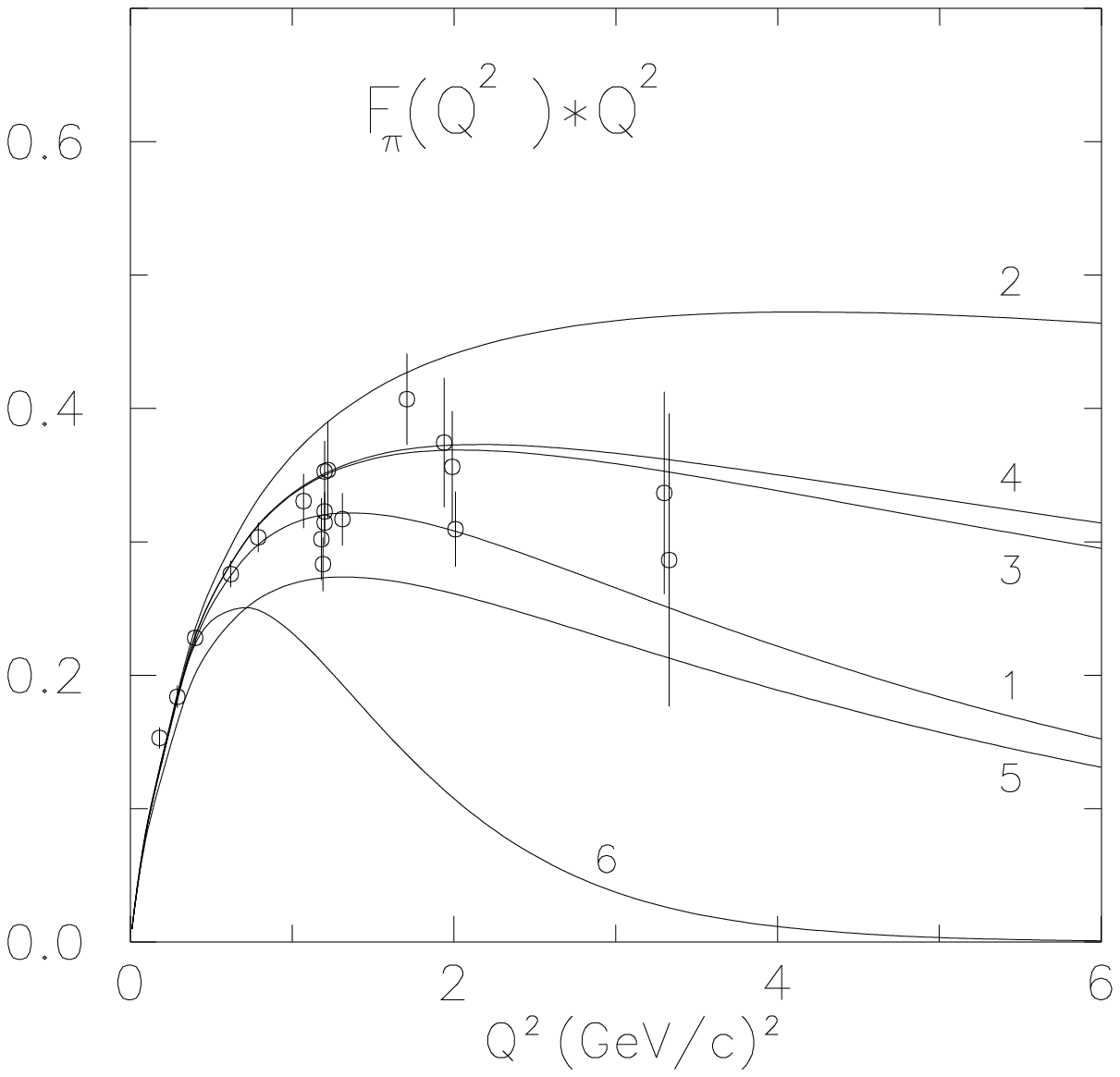}}}
\end{picture}
\caption{
Electromagnetic pion form factor, $Q^2 F_{\pi}(Q^2)$,
at high momentum transfer.
$M_u = M_d = 0.25$ $GeV/c$.
1 -- harmonic
 oscillator wave function Eq.(2), $b= $ 0.207 GeV;
2 -- power-law
wave function Eq.(3), $n = $ 2,  $b= $ 0.274 GeV; 
3 -- power-law wave
function Eq.(3), $n = $ 3,  $b= $ 0.388 GeV; 
4 -- wave function Eq.(4) from
model with linear confinement [6],  $a =$
0.0183 $\hbox{GeV}^2$, $b= $  0.7867.
5 -- harmonic
oscillator wave function Eq.(2), $b= $ 0.207 GeV ;
without spin rotation.
6 -- harmonic
oscillator wave function Eq.(2), $b= $ 0.207 GeV ;
nonrelativistic limit of Eq.(1).
Experimental data are taken from [7],[8].
}
\end{figure}

\newpage

\begin{figure}[p]
\begin{picture}(18,12)(-1,0)
\put(4,1){\makebox[8cm]{\epsfbox{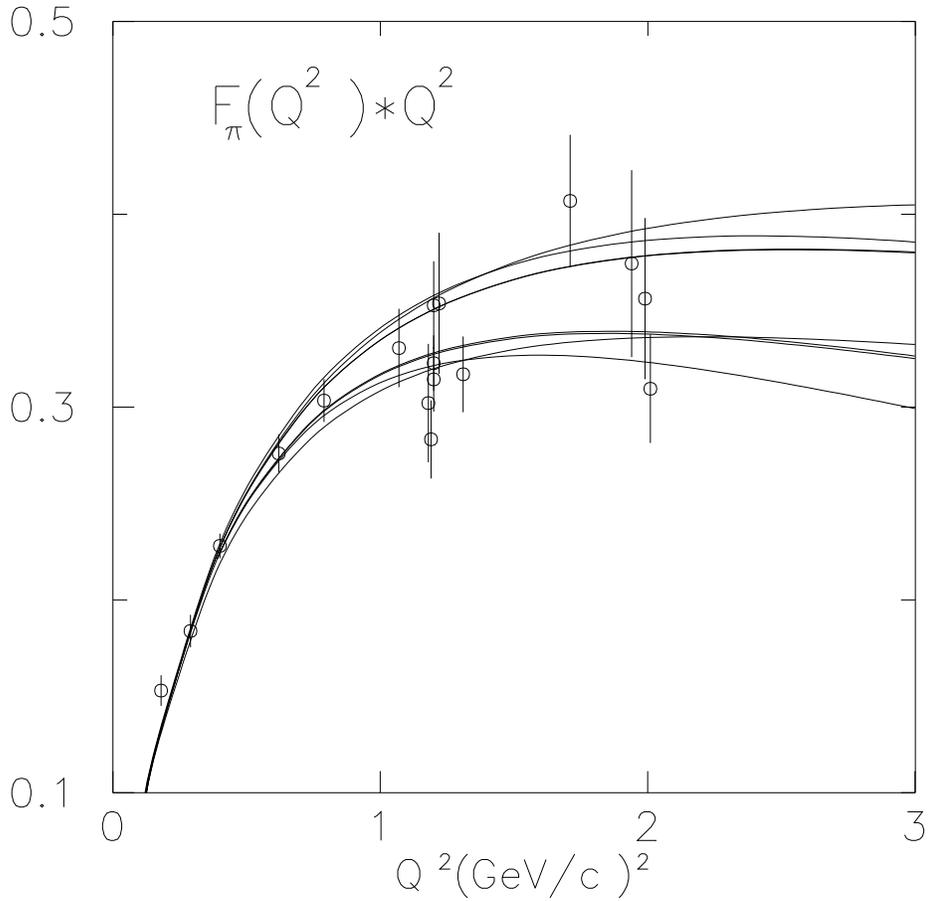}}}
\end{picture}
\caption{
$\pi$--meson form factor 
for different values
of quark anomalous magnetic moments and different wave 
functions.
Two sets of curves: the upper set is for the sum of 
quarks anomalous magnetic moments (-- 0.1), the lower --
for (+ 0.1).
If the quark structure is taken into account the
following effects are found:
1. Weak dependence on the wave function choice.
2. Strong dependence on the value of quark anomalous magnetic
moments.
}
\end{figure}

\begin{figure}[p]
\begin{picture}(18,12)(-1,0)
\put(4,1){\makebox[8cm]{\epsfbox{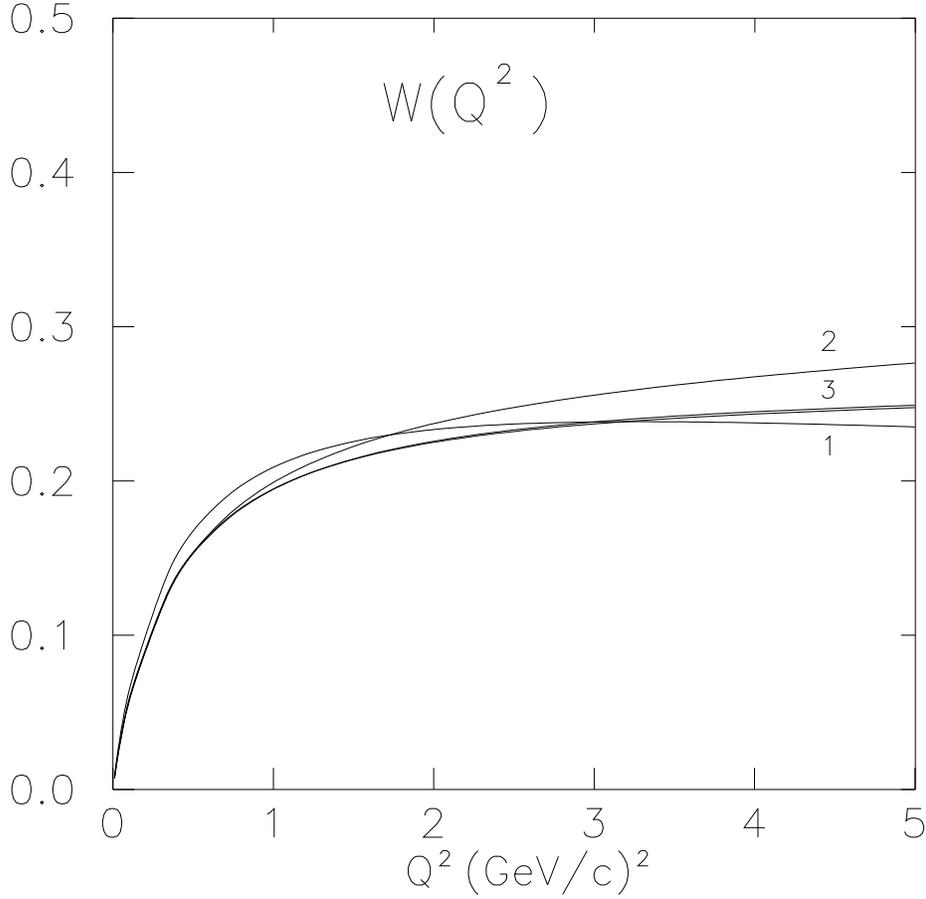}}}
\end{picture}
\caption{
The relative contribution of the effect of Wigner spin rotation
to $\pi$--meson form factor, 
$
W(Q^2)=\frac{F_{R+SR}(Q^2) - F_{R}(Q^2)}{F_{R+SR}(Q^2)}\>.
$
Here $F_{R+SR}(Q^2)$ -- pion form factor, spin rotation 
included,
$F_{R}(Q^2)$ -- without spin rotation.
$M_u = M_d = $0.25 GeV.
$<r_q^2> \simeq \frac{\hbox{0.3}}{M_q^2}$ = 0.187 Fm$^2$
(see {\it e.g.} [4]).
The same line code as in Fig.1 is used.
1 --  HO, $b=$0.350;
2 -- PL $n=$2, $b=$0.396;
3 -- PL $n=$3, $b=$0.571;
4 -- model with linear confinement $a=$0.083, $b=$0.7867.
The curves 3 and 4 practically coincide.
}
\end{figure}

\begin{figure}[p]
\begin{picture}(18,12)(-1,0)
\put(4,1){\makebox[8cm]{\epsfbox{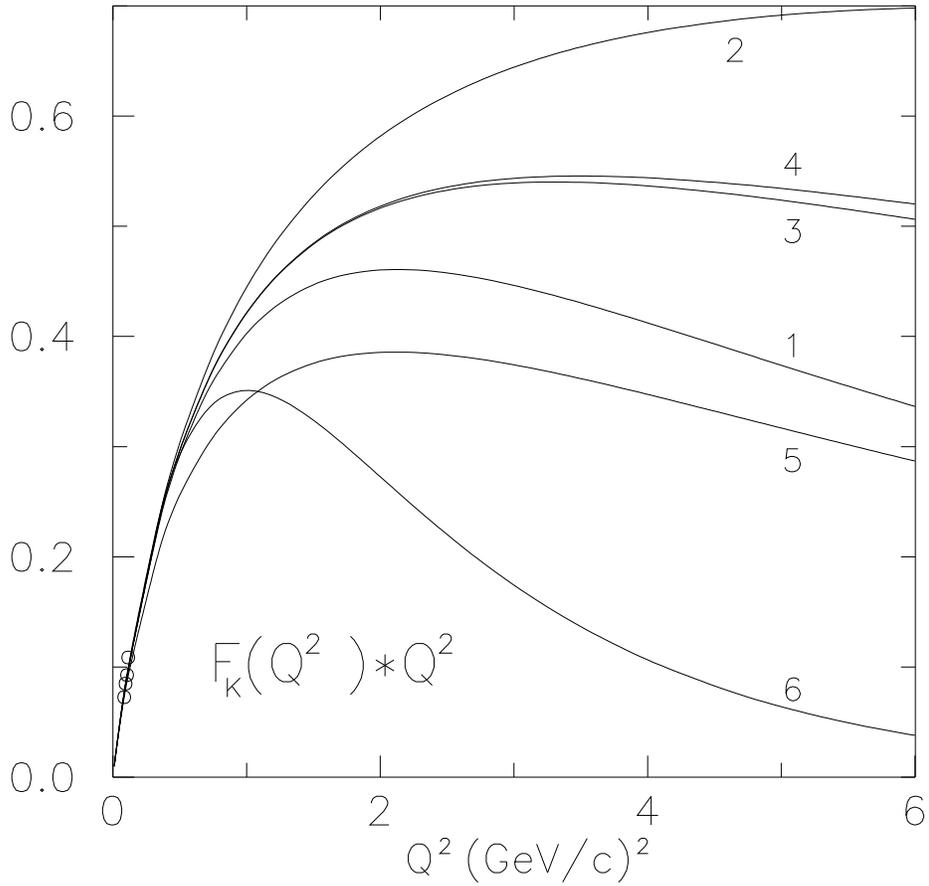}}}
\end{picture}
\caption{
Electromagnetic kaon form factor at high momentum transfer.
The same line code as in Fig.1 is used.
$M_s =$ 0.35 GeV, 
1 -- $b = $ 0.255 GeV,
2 -- $b =$ 0.339 GeV,
3 -- $b =$ 0.480 GeV,
4 -- $a= $ 0.0318 $\hbox{GeV}^2$, $b= $  0.7867.
Experimental data are taken from [9].
}
\end{figure}

\begin{figure}[p]
\begin{picture}(18,14)(-1,-1)
\put(4,1){\makebox[8cm]{\epsfbox{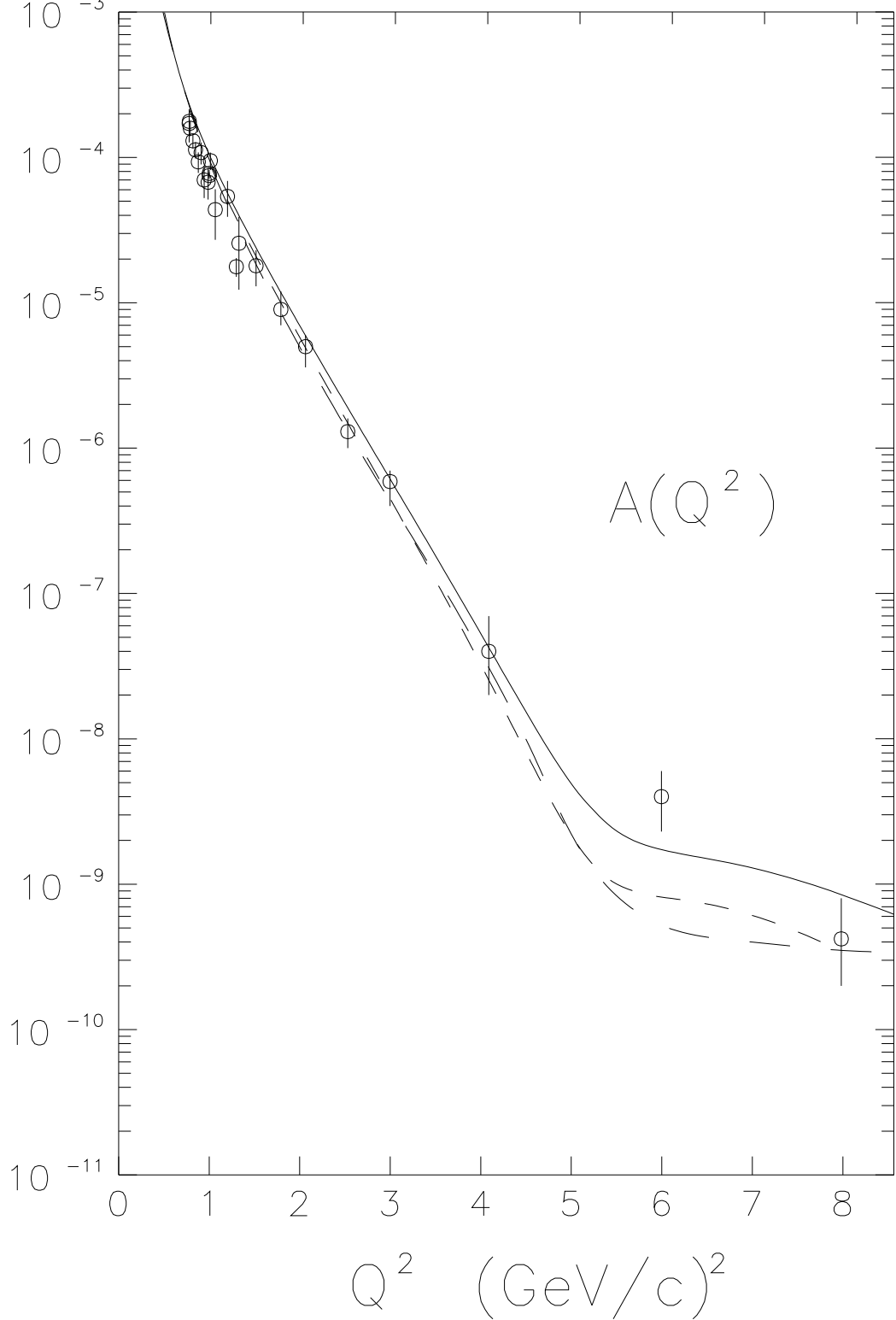}}}
\end{picture}
\caption{
Deuteron structure function $A(Q^2)$
for Paris wave functions from [10] and
nucleon form factors from [11]. Solid line is our relativistic
result, dashed--dot line is relativistic result from [12],
dashed line is nonrelativistic calculation.
Data are from [13].
}
\end{figure}

\begin{figure}[p]
\begin{picture}(18,12)(-1,0)
\put(4,2){\makebox[8cm]{\epsfbox{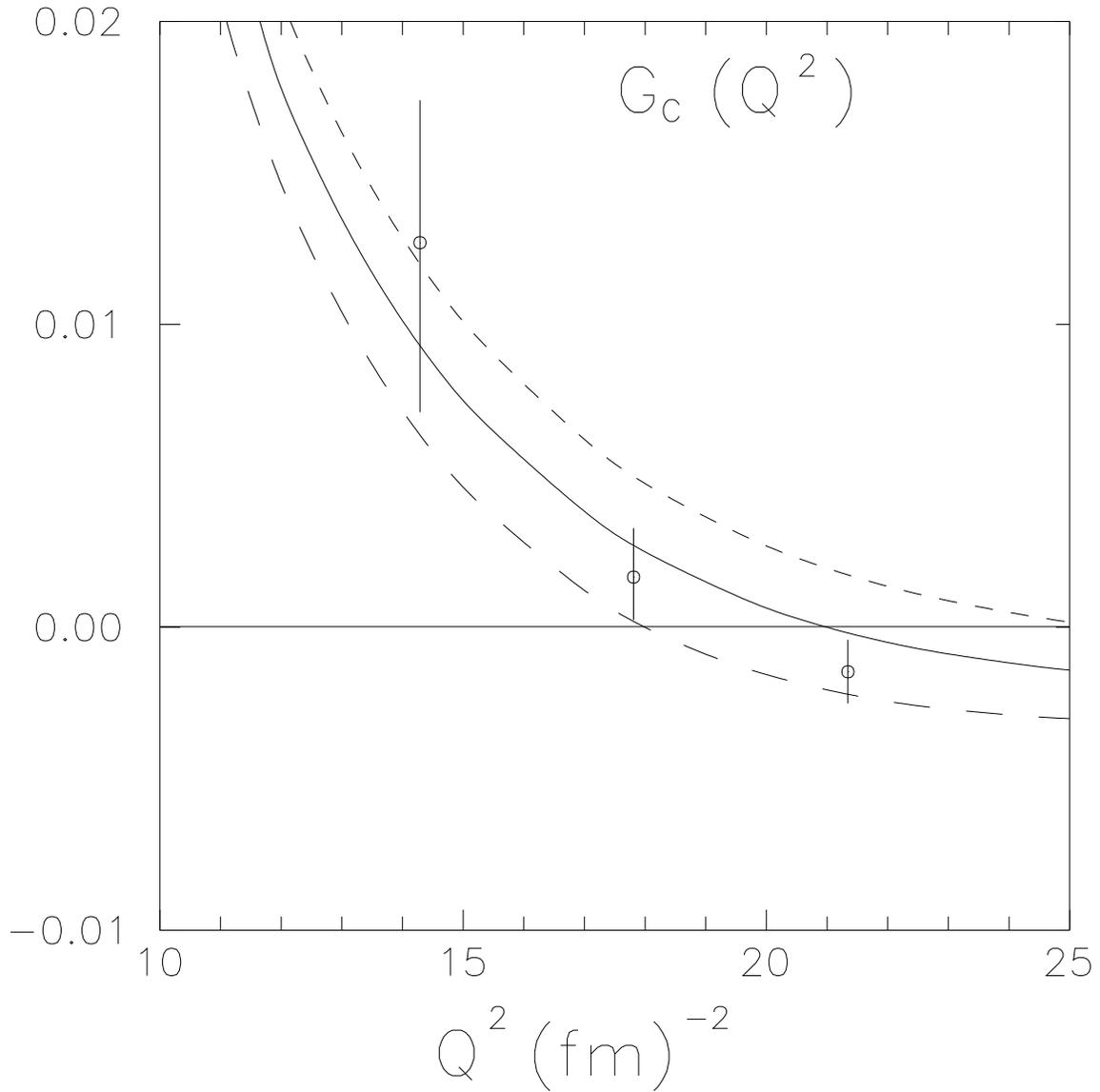}}}
\end{picture}
\caption{
The deuteron charge form factor $G_c(Q^2)$ 
near the point of zero value.
Solid line -- Paris wave functions from [10],
short--dashed line -- Bonn (R) [14],
long--dashed
line -- dispersion wave functions from [15].
For nucleon form factor the dipole fit is used.
}
\end{figure}

\begin{figure}[p]
\begin{picture}(18,9)(-1,0)
\put(4,2){\makebox[8cm]{\epsfbox{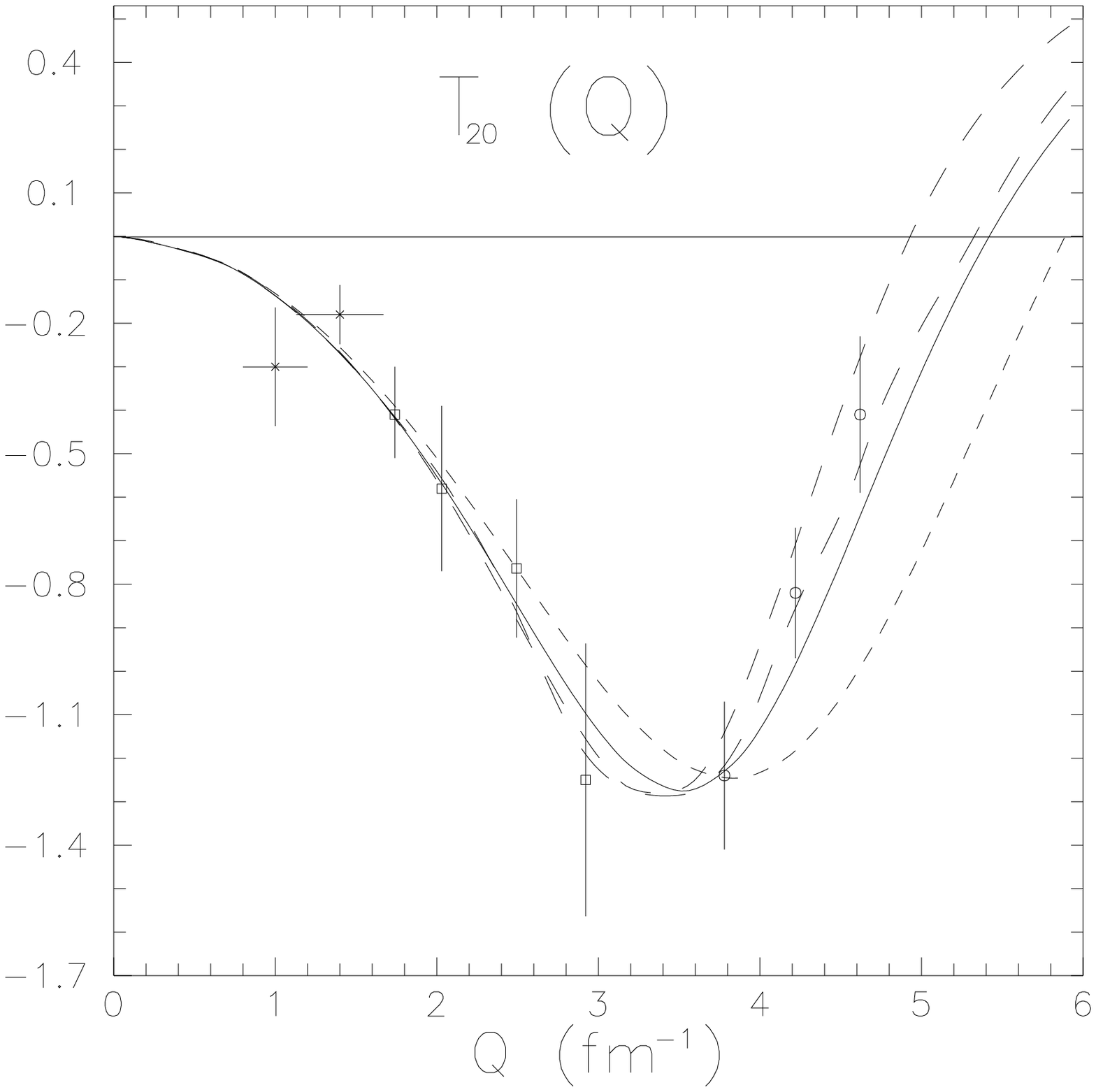}}}
\end{picture}
\caption{
Deuteron tensor polarization $T_{20}(Q^2)$.
$\vartheta = 70^{\mbox o}$.
Solid line -- Paris wave functions from [10],
short--dashed line -- Bonn (R) [14],
long--dashed
line -- dispersion wave functions from [15],
dashed--dot line is calculation in the Paris model from
[12]. 
Data are from [16].
}
\end{figure}


\begin{thebibliography}{99}

\bibitem{KeP91}
B.D.Keister and W.N.Polyzou,
Adv.Nucl.Phys. {\bf 20} (1991) 225.

\bibitem{KrT96}
Krutov~A.F. and Troitsky~V.E. J.
Phys. G: Nucl. Part. Phys. {\bf 19} (1993) L127;
Balandina~E.V., Krutov~A.F. and Troitsky~V.E.
{\it Theor. Mat. Fiz.} {\bf 103} (1995) 41
(Engl. transl. {\it Theor.
Math. Phys.} {\bf 103} (1995) 381);
J.Phys.G: Nucl. Part. Phys. {\bf 22}
(1996) 1585.

\bibitem{Lev95}
F.Lev 1995 E-print hep-ph/9505373.

\bibitem{CaG95}
F.Cardarelli, I.M.Grach, I.M.Narodetskii, E.Pace, G.Salme\'e
and S.Simula nucl-th/9507038.

\bibitem{Sch94} F.Schlumpf Phys.Rev. D {\bf 50} (1994) 6895.

\bibitem{Tez91} Tezuka~H. 1991 J. Phys. A: Math. Gen.
{\bf 24} (1991) 5267.


\bibitem{Ame84}
S.R.~Amendolia et al., Phys.Lett. {\bf B 146} (1984)116.


\bibitem{Beb78}
C.J.~Bebek et al., Phys.Rev. {\bf D 17}(1978)1693.

\bibitem{Ame86}
S.R.Amendolia {\it et al.}, Phys.Lett. {\bf 178B} (1986) 435.

\bibitem{Lac91}
M.Lacomb {\it et al.}, Phys.Lett. {\bf 101B} (1991) 139.

\bibitem{GaK85}
M.Gari and W.Kr\"umpelmann, Z.Phys. {\bf 322} (1985) 689.

\bibitem{ChC88}
P.L.Chung, F.Coester, B.D.Keister and W.N.Polyzou, Phys.Rev.
{\bf 37} (1988) 2000.

\bibitem{Arn75}
R.G.Arnold {\it et al.}, Phys.Rev.Lett. {\bf 35} (1975) 776;
G.G.Simon {\it et al.}, Nucl.Phys. {\bf A364} (1981) 285;
N.de Botton, Nucl.Phys. {\bf A374} (1982) 143.

\bibitem{Mac87}
R.Machleidt {\it et al.}, Phys.Rep. {\bf 149} (1987) 1.

\bibitem{MuT81}
V.M.Muzafarov and V.E.Troitsky, Yad.Fiz. {\bf 33} (1981) 2040.

\bibitem{Sch84}
M.E.Schulze {\it et al.}, Phys.Rev.Lett. {\bf 52} (1984) 597;
V.E.Dmitriev {\it et al.}, Phys.Lett. {\bf B157} (1985) 143;
B.B.Voitsekhovskii {\it et al.}, Pis'ma JETP {\bf 43} (1986)
567; R.G.Gilman {\it et al.}, Phys.Rev.Lett. {\bf 65} (1990)
1733; I.The {\it et al.}, Phys.Rev.Lett. {\bf 67} (1990) 173.
\end{thebibliography}
\end{document}